\documentclass[nofootinbib,showkeys,superscriptaddress,preprintnumbers,floatfix]{revtex4-1}

\usepackage{dynlearn}
\usepackage{enumitem}
\usepackage{geometry}
\usepackage{braket}

\newcommand{\CO}{\mathrm{CO}}
\begin{document}

\def\ourTitle{Nonequilibrium Thermodynamics in Measuring Carbon Footprints:\\
Disentangling Structure and Artifact in\\
Input-Output Accounting}

\def\ourAbstract{Multiregional input-output (MRIO) tables, in conjunction with Leontief analysis,
are widely-used to assess the geographical distribution of carbon emissions and
the economic activities that cause them. Majorization, a tool originating in
economics that has found utility in statistical mechanics, can provide insight
into how Leontief analysis links disparities in emissions with global income
inequality. We examine Leontief analysis as a model, drawing out similarities
with modern nonequilibrium statistical mechanics. Paralleling the physical
concept of thermo-majorization, we define the concept of eco-majorization and
show it is a sufficient condition to determine the directionality of embodied
emission flows. Surprisingly, relatively small trade deficits and a
geographically heterogeneous emissions-per-dollar ratio greatly increases the
appearance of eco-majorization, regardless of any further content in the MRIO
tables used. Our results are bolstered by a statistical analysis of null models
of MRIO tables, based on data provided by the Global Trade Aggregation Project.
}

\def\ourKeywords{majorization, nonequilibrium thermodynamics,
input-output analysis, carbon footprint,
ecologically unequal exchange
}

\hypersetup{
  pdfauthor={Samuel P. Loomis},
  pdftitle={\ourTitle},
  pdfsubject={\ourAbstract},
  pdfkeywords={\ourKeywords},
  pdfproducer={},
  pdfcreator={}
}

\author{Samuel P. Loomis\footnote{Corresponding author.}}
\email{sloomis@ucdavis.edu}
\affiliation{Complexity Sciences Center and Physics Department,
University of California at Davis, One Shields Avenue, Davis, CA 95616}

\author{Mark Cooper}
\email{mhcooper@ucdavis.edu}
\affiliation{Department of Human Ecology,
University of California at Davis, One Shields Avenue, Davis, CA 95616}

\author{James P. Crutchfield}
\email{chaos@ucdavis.edu}
\affiliation{Complexity Sciences Center and Physics Department,
University of California at Davis, One Shields Avenue, Davis, CA 95616}

\date{\today}
\bibliographystyle{unsrt}

\title{\ourTitle}

\begin{abstract}
\ourAbstract
\end{abstract}

\keywords{\ourKeywords\\
DOI: \url{XX.XXXX/....}}

\preprint{\arxiv{2106.03948}}

\title{\ourTitle}

\date{\today}

\maketitle

\setstretch{1.1}


\newcommand{\kB}{k_\text{B}}

\section{Introduction}
As our planet faces environmental catastrophe of unprecedented scope, it has
become necessary to address human impacts on the climate and global ecosystem
through multilateral action by the world's governments. Warnings have been
raised about the pitfalls of too short-sighted a response. For instance,
treaties that only address pollution at the point of production may effectively
outsource carbon-intensive activities from signatory nations to nonsignatories,
a process known as \emph{carbon leakage}
\cite{peters_growth_2011,davis_supply_2011}. Instead, a holistic response that
accounts for the multifaceted social relations driving environmental impacts is
required \cite{hornborg_unequal_2003, rice_ecological_2007, smith_trade_2012,
jorgenson_sociology_2012-1, bergmann_bound_2013}. Acquiring the data needed to
make such holistic assessments, however, is a challenge in its own right.

Multiregional input-output (MRIO) tables provide data on the monetary
transactions between national-level industries, both within a nation's borders
and across them \cite{leontief_economy_1991,leontief_essays_1966,
wiedmann_editorial_2009, kitzes_introduction_2013,
schaffartzik_environmentally_2014}. These can be used to construct models of
the interconnected global economy. MRIO tables can be ecologically extended
(called EE-MRIO tables) by adding local data on the environmental impacts that
arise as byproducts of production. 

Leontief analysis is a method frequently paired with EE-MRIO tables to
attribute production-level impacts to the (potentially distant) activities that
they support---typically consumption \cite{krausmann_global_2008,
peters_production-based_2008, erb_embodied_2009, moran_trading_2009,
davis_consumption-based_2010, wiedmann_carbon_2010,
peters_growth_2011,davis_supply_2011, bergmann_bound_2013, moran_does_2013,
yu_tele-connecting_2013, alsamawi_employment_2014, prell_economic_2014,
simas_bad_2014, dorninger_can_2015, liu_carbon_2015, bergmann_land_2016,
oita_substantial_2016, oosterhaven_basic_2019}. These attributed impacts are
said to be \emph{embodied} in the consumed product. The flows of embodied
impacts computed from EE-MRIO tables have been utilized in policy analysis by
global and national government institutions
\cite{nl_report_2012,un_report_2016,aus_report_2016,world_dev_report}. Leontief
analysis makes key assumptions, however, whose accuracy has been brought into
question, particularly when applied to existing MRIO tables
\cite{kitzes_introduction_2013,
schaffartzik_environmentally_2014,schaffartzik_raw_2015}. The following
explores the \emph{consequences} of these assumptions, through the lens of
statistical mechanics.

In particular, we find that when certain reasonable conditions hold on an
EE-MRIO dataset---namely, relatively small trade deficits among nations and
geographically heterogeneous impact intensities---the directionality of
embodied impact flows to and from extremal regions is heavily influenced by the
impact intensities. Notably, the MRIO tables themselves have only a secondary
effect. We call the phenomenon mediating this \emph{eco-majorization}.

Our analysis of eco-majorization relies on the general theory of majorization
and Lorenz curves \cite{marshall_inequalities_2011} which have found wide
application in economic and social analysis
\cite{lorenz_methods_1905,veinott_least_1971}, statistical decision theory
\cite{blackwell_comparison_1951,blackwell_equivalent_1953}, and statistical
physics \cite{janzing_thermodynamic_2000, Horo09a, brandao_resource_2013,
horodecki_fundamental_2013, brandao_second_2015,Lost18a}. Majorization and Lorenz
curves are a means of characterizing the differences between two distributions
without reducing those differences to a single parameter. The intuition of
majorization has a natural foothold in the assumptions of Leontief analysis and
EE-MRIO tables. Due to this, heterogeneities in impact intensities drive
embodied flows in a manner directly analogous to physical diffusion.

Disentangling the results of a mode of analysis from mathematical artifacts
that arise from the assumptions entailed is a difficult and often overlooked
practice when working with complex data. For this reason, the use of
specialized null models in network science has become increasingly popular
\cite{whitehead_investigating_1995, serrano_correlations_2006,
ansmann_constrained_2011, prettejohn_methods_2011, rankin_role_2016,
farine_guide_2017}. The null models are used to randomly generate networks with
special constraints designed to replicate the structural assumptions of a
dataset while otherwise reducing structural biases via random connections.
Following this, we use null models specifically constructed to address the
structures of EE-MRIO datasets, providing numerical confirmation of how
majorization mediates the relationship between the assumptions of Leontief
analysis and the embodied flows it detects.

\Cref{sec:background} covers the relevant mathematics of EE-MRIO tables and
Leontief analysis, the
construction of our null models, and 
the necessary details for using
majorization. \Cref{sec:results} shows how our symmetrically-constructed
null model of trade recovers asymmetric embodied flows quite regularly. To
explain this, we define eco-majorization, demonstrate how it couples the impact
intensities to flows of embodied resources, and explain why (under the
aforementioned conditions) Leontief analysis is structurally biased towards
eco-majorization. Further analysis of the null model shows that relaxing our
conditions mitigates majorization's effects. \Cref{sec:discussion} closes
discussing the implications of these results for analyzing EE-MRIO tables and
environmental accounting.

\section{Background and Methods}
\label{sec:background}

\subsection{Input-output analysis}
\label{sec:input-output}

The following describes the basic components of EE-MRIO analysis, focusing on
aspects relevant to the developments in Sec. \ref{sec:results}. Useful reviews
of input-output methods are found in Refs.
\cite{schaffartzik_environmentally_2014,kitzes_introduction_2013,
oosterhaven_basic_2019}. Additionally, we present a somewhat more detailed
description of these methods in App. \ref{app:io-analysis}.

\begin{figure}[t]
\centering
\includegraphics[width=0.6\columnwidth]{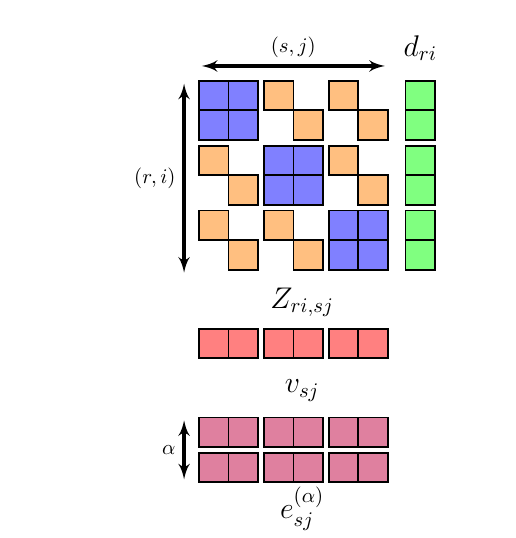}
\caption{Block-matrix structure of a typical EE-MRIO table with
	$|\mathcal{I}|=2$ and $|\mathcal{R}|=3$. Blue and orange blocks are
	arranged into the single block-matrix $Z_{ri,sj}$, representing the value
	of all inter-industry transactions. Red blocks correspond to the
	value-added vector $v_{ri}$, consisting of all returns to wages, profits,
	and rent. And, the green blocks correspond to the final demand vector
	$d_{ri}$, consisting of all expenditures by consumers, governments, and
	investors. Pink blocks represent two separate ecological impact
	distributions $e^{(\alpha)}_{ri}$.
	}
\label{fig:io}
\end{figure}

A MRIO table consists of a set of regions $\mathcal{R}$ and a set of industrial
sectors $\mathcal{I}$. The full global set of sectors is given by
$\mathcal{R}\times\mathcal{I}$, with pair $(r,i)\in
\mathcal{R}\times\mathcal{I}$ corresponding to industry $i$ in region $r$.  An
input-output table over these industries is characterized by three pieces: the
\emph{value-added} block-vector $v_{ri}$, describing the incomes paid by
industry $i$ to labor, land, and capital in the region $r$; the \emph{demand}
block vector $d_{ri}$, describing the value purchased from industry $i$ by
private or government spending and investment; and the inter-industrial block
matrix $Z_{ri,sj}$ that describes the flow of money from industry $j$ in region
$s$ to industry $i$ in region $r$. These quantities are assumed to be measured
over a fixed time interval. By focusing on monetary flows, we are implicitly
assuming that the necessary material requirements for each industry and demand
are met, and that supply equals demand overall. The effects of relaxing this
assumption are beyond the scope of the present paper.
  
In a block matrix each row and column indicates a pair $(r,i)$, so that the
first $|\mathcal{I}|$ indices correspond to all the industries in one region,
and so on; Fig. \ref{fig:io} gives a visual aid. Our notation indicates this
through use of commas: $Z_{ri,sj}$ is a compact way of denoting the matrix
element $Z_{r\times |\mathcal{I}| + i,s\times |\mathcal{I}|+j}$, and similarly
for $v_{ri}$ and $d_{ri}$.

Additionally, simplifying
constructs are often used in collecting and apportioning the data. For
instance, in the dataset GTAP 8 (discussed shortly), $\mathcal{I}$ contains a
duplicate of each industry, one for managing imports and the other for domestic
industries. Cross-regional trade from $r$ to $s$ can only go from the domestic
copy of industry $i$ in $r$ to the import copy of industry $i$ in $s$. That is,
cross-regional trade is not treated as cross-industry. This is represented
visually by the diagonal structure of off-diagonal blocks in Fig. \ref{fig:io}.

MRIO table components are constrained by the following identity which demands
all accounts balance. That is, for each industry, all money outlaid on
materials and incomes equals the total value of output purchased. The total
throughput is denoted by the block-vector $z_{ri}$:
\begin{equation}
\label{eq:balance}
  z_{ri}  := \underbrace{
        v_{ri} 
      + \sum_{s,j} Z_{sj,ri}
      }_{
        \text{Outlays}
      } 
       = \underbrace{
        \sum_{sj} Z_{ri,sj}
        + d_{ri}
      }_{
        \text{Outputs}
	}
	~.
\end{equation}
A consequence is that global income $y := \sum_{ri}v_{ri}$ and global spending
$\sum_{ri} d_{ri}$ are equal. Whereas, \emph{regional} income $\hat{y}_r :=
\sum_{i}v_{ri}$ and spending $\hat{x}_r := \sum_i d_{ri}$ are \emph{not}
necessarily equal. We refer to their difference $\hat{x}_r-\hat{y}_r$ as the
\emph{regional deficit}.

The technical coefficients $C_{ri,sj}:= Z_{ri,sj}/z_{sj}$ express what
proportion of outlays are committed to specific industries. Clearly,
$Z_{ri,sj}= z_{sj} C_{ri,sj}$. Given this, the balance identity
\cref{eq:balance} may be expressed as a linear-algebraic equation $\mathbf{z} =
\mathbf{C}\mathbf{z}+\mathbf{d}$, with matrix multiplication between
block-matrix $\mathbf{C}$ and block-vector $\mathbf{z}$. Solving this equation
for $\mathbf{z}$ gives:
\begin{align}
\label{eq:leontief-solution}
  \mathbf{z} = \left(\mathbf{I}-\mathbf{C}\right)^{-1} \mathbf{d}
  ~,
\end{align}
where $\mathbf{I}$ is the identity matrix and one uses the matrix inverse.
Eq. \ref{eq:leontief-solution} expresses $\mathbf{z}$ in terms of a different
set of variables than those usedd in Eq. \ref{eq:balance}, though both equations
are equally true. This new expression, however, is useful in decomposing
$\mathbf{z}$ in terms of the demands which generate its value.
Written more explicitly, we have:
\begin{align}
\label{eq:leontief-expanded}
  z_{ri} = \sum_{sj} \left[\left(\mathbf{I}-\mathbf{C}\right)^{-1} \right]_{ri,sj}d_{sj}
  ~.
\end{align}
This expresses the total output as a column-sum of the matrix
$\left(\mathbf{I}-\mathbf{C}\right)^{-1} \mathbf{D}$. The sum allows us to
break sector $i$'s total output into parts, each attributed to a particular
final region of demand $s$. The proportions of this decomposition are given by
the attribution matrix $\mathbf{A}$:
\begin{align}
\label{eq:attribution}
  A_{ri,s} := \frac{\sum_{j}\left[\left(\mathbf{I}-\mathbf{C}\right)^{-1} \right]_{ri,sj}d_{sj}}{z_{ri}}
  ~,
\end{align}
which characterizes what portion of the throughput of sector $(r,i)$ originated
as demand in region $s$. Determining these attributions is called
\emph{Leontief analysis} after its originator
\cite{leontief_economy_1991,leontief_essays_1966}.\footnote{The analysis may be
reversed to attribute outputs to factors $\mathcal{Y}$ rather than to final
demands $\mathcal{D}$. While not explicitly considered here, the results
derived for demand-based accounting apply symmetrically to factor-based
accounting.}

The application of Leontief analysis which occupies us is its use in
attributing the \emph{ecological impacts} of particular industries to the
parties whose demand stimulates the impact. This is accomplished through the
augmentation of an MRIO table with \emph{environmental extensions}. These are a
series of block-vectors $e^{(\alpha)}_{ri}$ (the series being indexed by
$\alpha$) each of which characterizes what quantity of ecological impact
$\alpha$ may be directly attributed to the industry $i$ in region $r$. For
instance, if $\alpha$ is $\mathrm{CO}_2$, then $e^{(\alpha)}_{ri}$ is the
total megatons of $\mathrm{CO}_2$ released by the activity in sector $(r,i)$.
If $\alpha$ is labor-time, then $e^{(\alpha)}_{ri}$ is the amount of
human-hours utilized by sector $(r,i)$.

We denote the \emph{regional direct impact} by $\hat{e}^{(\alpha)}_r := \sum_i
e^{(\alpha)}_{ri}$ and the \emph{total impact} by $E^{(\alpha)} := \sum_r
\hat{e}^{(\alpha)}_r$. Using the attribution matrix, we then define the
\emph{attributed impact vector} as:
\begin{align}
\label{eq:attr-def}
  \hat{a}_{s}^{(\alpha)} := \sum_{ri} e^{(\alpha)}_{ri} A_{ri,s}
  ~.
\end{align}
This, in theory, characterizes the quantity of impact $\alpha$
for which the demand in region $s$ is responsible.

The validity of Eq. \ref{eq:attr-def} rests on two main assumptions
\cite{schaffartzik_environmentally_2014}:
\begin{enumerate}[label={(L-\arabic*)},align=left]
      \setlength{\topsep}{-5pt}
      \setlength{\itemsep}{-5pt}
      \setlength{\parsep}{-5pt}
\item \label{itm:L1} \textbf{Sectors produce homogeneous products}: Due to this, we
	do not reweight the technical coefficients to reflect differences between
	the purchasing sectors---they purchase the same item.
\item \label{itm:L2} \textbf{Sectoral products are homogeneously priced}: Every buyer
	pays the same unit price. This again allows using the technical
	coefficients without modification to reflect differences in the inputs
	required per dollar for different purchasers. 
\end{enumerate}
These have been particularly singled out because of their crucial role in
justifying the use of the single matrix $A_{ri,s}$ to compute all attribution
vectors. 
We will return to the assumptions later to discuss how this constraint impacts
the results of Leontief analysis.

For now, we mention two important reflections of these assumptions in the
results above. First is the fact that $\sum_s A_{ri,s}=1$ for all regions $r$
and industries $i$. Additionally, because $(\mathbf{I}-\mathbf{C})^{-1}$ is a
positive matrix \cite{suh_2007}, $A_{ri,s}$ is also positive. These two facts give the matrix
$A_{ri,s}$ the property of \emph{stochasticity}. That the direct impacts and
attributed impacts can be related by a single stochastic matrix is a reflection
of the homogeneity in each of these assumptions, and is the central factor at
play in our results.

Second, income itself can be treated as an impact. The value-added vector, when
passed through the attribution matrix, returns the regional spending vector:
\begin{align}
  \hat{x}_{s} = \sum_{ri} v_{ri} A_{ri,s}\ ~,
\label{eq:attr-income}
\end{align}
While App. \ref{app:io-analysis} explains the identity's mathematical origin,
conceptually it arises from assumption \ref{itm:L2}: Impacts are attributed to
demand in the exact same manner that incomes are attributed to spending. Both
the stochasticity of $A_{ri,s}$ and \cref{eq:attr-income} will be critically
important when applying majorization.

Finally, we define several miscellaneous terms that feature in our analysis.
Given impact $\alpha$, we define the \emph{$\alpha$-intensity}
$f^{(\alpha)}_{ri}$ of sector $(r,i)$ as the ratio of environmental impact to
economic impact. For instance, for $\mathrm{CO}_2$ this represents the
pollution released for each dollar of activity in sector $(r,i)$. The most
intuitive interpretation is seen in the case of labor-time, where the intensity
is a kind of inverse wage.

While there are many different precise definitions of intensity, for our needs
we define it for sectors and regions as:
\begin{align*}
  f^{(\alpha)}_{ri} := \frac{e^{(\alpha)}_{ri}/v_{ri}}{E^{(\alpha)}/Y} ~,
  \quad
  \hat{f}^{(\alpha)}_{r}  :=
  \frac{\hat{e}^{(\alpha)}_{r}/\hat{y}_{r}}{E^{(\alpha)}/Y}
\end{align*}
That is, we take the ratio of the impact $e^{(\alpha)}_{ri}$ 
to the \emph{value-added} 
$v_{ri}$ by
sector $(r,i)$. This directly relates the emission at a given stage of
production to the value added to the region by that production. (Not counted in
this is the economic input from other industries, as this value has already
been counted as value-added in another industry.) To remove dependence on the
units used, we normalize the ratio using the ratio of totals.

Embodied flows resulting from Leontief analysis are frequently quantified by
one or both of the following proxy measures \cite{erb_embodied_2009,
davis_consumption-based_2010, bergmann_bound_2013, moran_does_2013,
yu_tele-connecting_2013, alsamawi_employment_2014, simas_bad_2014,
dorninger_can_2015, liu_carbon_2015, bergmann_land_2016, oita_substantial_2016}:
\begin{enumerate}
      \setlength{\topsep}{-5pt}
      \setlength{\itemsep}{-5pt}
      \setlength{\parsep}{-5pt}
\item The net exports $\boldsymbol{\xi}^{(\alpha)}=(\xi^{(\alpha)}_r)$ 
of attributed impact, as a share of total global impact:
  \begin{align*}
    \xi^{(\alpha)}_r 
    = \frac{\hat{e}^{(\alpha)}_r - \hat{a}^{(\alpha)}_r}{\sum_r \hat{e}^{(\alpha)}_r} ~.
  \end{align*}
\item The consumption/production ratio $\boldsymbol{\rho}^{(\alpha)}=(\rho^{(\alpha)}_r)$ 
  of attributed impact to direct impact:
  \begin{align*}
    \rho^{(\alpha)}_r
    = \frac{\hat{a}^{(\alpha)}_r}{\hat{e}^{(\alpha)}_r}
  ~.
  \end{align*}
\end{enumerate}
Naturally, these are closely related, as they ultimately express the
relationship between the relative sizes of produced impacts and consumed
impacts.

\subsection{Null models and GTAP 8}
\label{sec:gtap}

Null models, also \emph{configuration models}, are a popular tool in the study
of complex networks. They are widely applied to social
\cite{whitehead_investigating_1995,kojaku_core-periphery_2018}, animal
\cite{rankin_role_2016,farine_guide_2017}, and biological
\cite{prettejohn_methods_2011} networks to separate-out structures detected in
empirical networks from those engendered by methodological assumptions
\cite{serrano_correlations_2006,ansmann_constrained_2011}. Often particular
aspects of the networks---such as degree distribution---are held constant while
all remaining aspects are randomized.

The following constructs a null model of global trade that maintains similar
technical coefficients between industries, but ``social coefficients''---those
determining the relations between nations, dependency on imports, proportion
paid to factors, and so on---are entirely randomized. This way, structures that
consistently arise from Leontief analysis of networks drawn from this model
cannot be attributed to social relations. They are, rather, artifacts of the
assumptions of Leontief analysis.

The Global Trade Aggregation Project (GTAP) \cite{narayanan_g_global_2012}
offers an extensive collection of transaction tables over a large number of
regions and sectors. GTAP
has been used as the EE-MRIO source data in many studies of embodied
carbon emissions and other impacts \cite{davis_consumption-based_2010,
bergmann_bound_2013,
yu_tele-connecting_2013, prell_economic_2014, 
bergmann_land_2016}.
We used GTAP 8 covering $134$ regions ($114$ of which are
countries), with $57$ industrial sectors within each region, as well as $5$
factor sectors (skilled and unskilled labor, capital, land, and natural
resources), the standard $3$ final demand sectors, and further sectors that
fall outside the scope of our analysis. The data on these sectors was used to
construct a multiregional input-output table. GTAP 8 comes directly with
environmental extensions for carbon and energy use, and a number of satellite
datasets exist providing further extensions. As an additional point of
comparison, we used the satellite GMig2 \cite{walmsley_global_2007} that
provides information on labor inputs in human-years.

The carbon and labor data provided by GTAP 8 and GMig2 allowed us to compute
the carbon and labor intensities, $\hat{\mathbf{f}}^{(\CO_2)}$ and
$\hat{\mathbf{f}}^{(\mathrm{L})}$, respectively, over $17$ \emph{megaregions}
formed by aggregating the $134$ GTAP standard regions.

We then constructed a null model for generating trade datasets over a reduced
MRIO system with $4$ factors, $16$ industrial sectors, and $17$ regions
(aggregated from the GTAP sectors and regions). The null model has two
parameters: a scalar $\zeta_X$ and a vector
$\zeta_{C,i}$ taking values over industrial sectors. It
makes liberal use of Dirichlet distributions as the source for drawing
randomly-generated stochastic matrices from which the input-output tables are
procedurally constructed.

Recall that a Dirichlet distribution $\mathrm{Dir}(\alpha_1,\dots,\alpha_K)$ is
defined on the simplex of probability vectors over $K$ elements.  The density
function of $\mathrm{Dir}(\alpha_1,\dots,\alpha_K)$ is given by:
\begin{align*}
  d(p_1,\dots,p_K) \propto \prod_{i=1}^K p_i^{\alpha_i-1}
  ~.
\end{align*}
Notably, when $\alpha_1,\dots,\alpha_K$ all equal $1$, the Dirichlet
distribution draws from all probability vectors with equal weight
(a uniform Dirichlet). Otherwise,
it tends to draw probability vectors whose weight distribution is similar to
that of the vector $(\alpha_1-1,\dots,\alpha_K-1)$, varying to a degree that is
inversely proportional to $\alpha_0:=\sum_{i=1}^K \alpha_i$
(a nonuniform Dirichlet).

The null model uses nonuniform Dirichlet distributions to draw the values of
the regional technical coefficients so that they are similar to the global
coefficients $\tilde{C}_{ij}^{(\mathrm{GTAP})}$ from GTAP 8 with a degree of
variation controlled by the parameter $\zeta_{C,i}$ for each industrial sector
$i$. All other coefficients---which we call \emph{social coefficients}---are
constructed from combining uniform Dirichlet distributions. Using the technical
and social coefficients, we can use another nonuniform Dirichlet to generate
the regional spending and income distributions so that the regional deficit is
small, controlled by parameter $\zeta_X$. Baseline values $\bar{\zeta}_{C,i}$
and $\bar{\zeta}_X$ are derived to replicate the variation among regional
technical coefficients and deficits in the GTAP 8 data. The precise
construction is described in Appendix \ref{app:null-model}.

Additionally, we built a second, simpler null model for producing environmental
extensions. It generates impact intensities, with control over the
heterogeneity of intensity across regions and sectors. The imaginary resource
that this impact represents is termed \emph{unobtainium} or simply
$\mathrm{U}$. The null model involves constructing a new parameter $\zeta_U$.
We sample \emph{unnormalized} intensities, denoted $\phi_{ri}^{(\mathrm{U})}$,
as:
\begin{align*}
  \phi_{ri}^{(\mathrm{U})} \sim \mathrm{Dir}(\alpha_{ri}=\zeta_U)
  ~.
\end{align*}
$\zeta_U$ is set to a low value---our baseline is
$\bar{\zeta}_U=0.05$---resulting in the Dirichlet sampling distributions with
high peakedness around randomly selected sectors and so assuring heterogeneity.
By increasing $\zeta_U$, we can reduce heterogeneity.  The normalized
intensities, for a given MRIO, are then determined as:
\begin{align*}
  f_{ri}^{(\mathrm{U})} =\frac{\phi_{ri}^{(\mathrm{U})}}{\sum_{s,j}v_{sj}\phi_{sj}^{(\mathrm{U})}}
  ~.
\end{align*}
Combining both null models allows generating a wide sampling of EE-MRIO tables with random social coefficients, while controlling for technical coefficients, regional deficits, and impact heterogeneity. These last two attributes, in particular, are important for majorization, as we will show.

\begin{figure*}[t]
\centering
\includegraphics[width=0.32\textwidth]{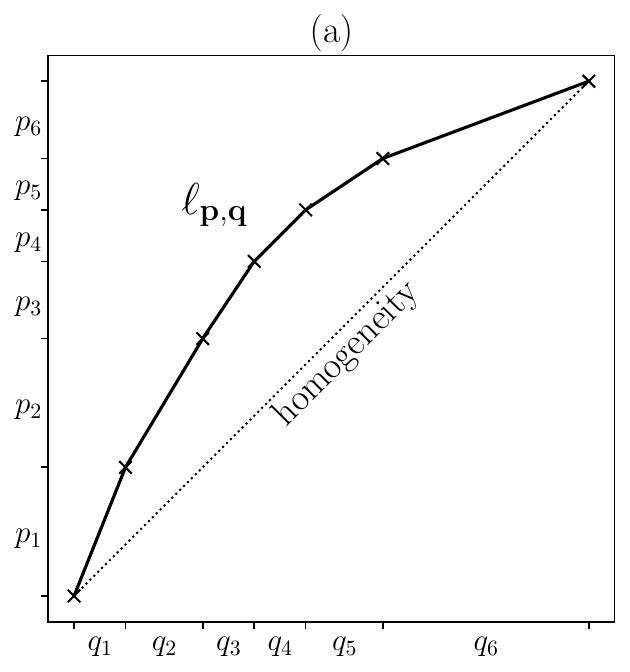}
\includegraphics[width=0.32\textwidth]{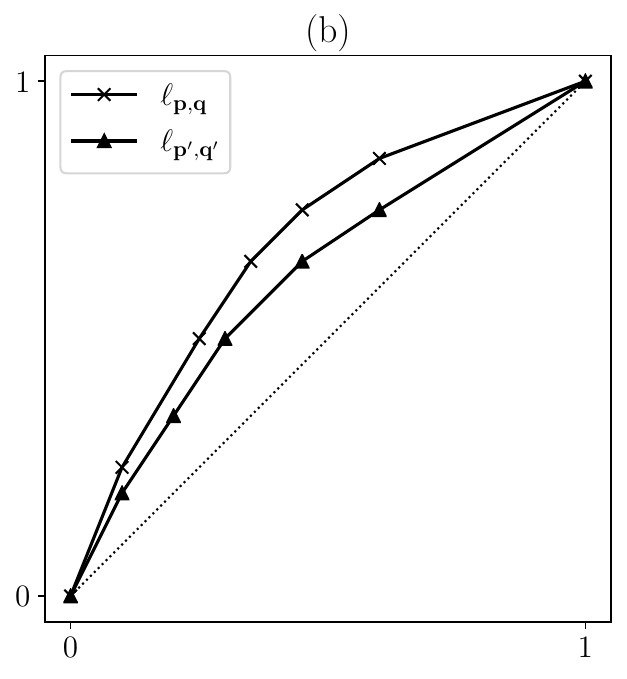}
\includegraphics[width=0.32\textwidth]{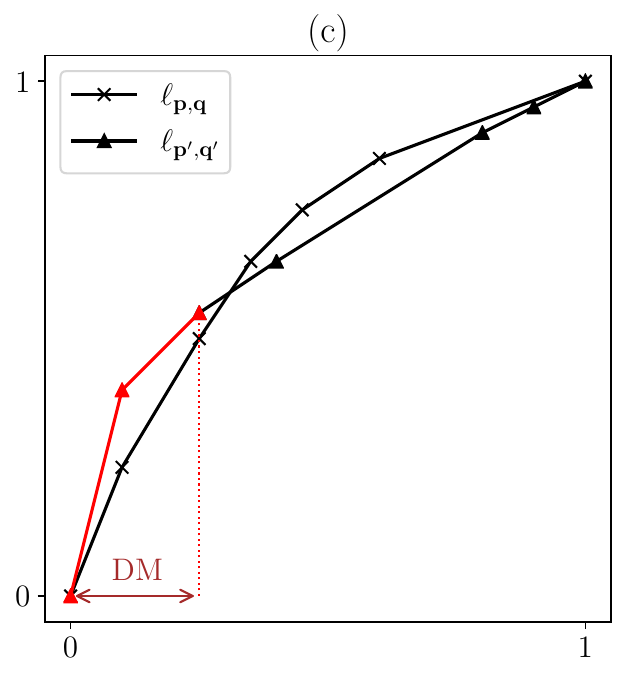}
\caption{Majorization Primer: \emph{(a)} Example a Lorenz curve for a pair of
	distributions $(\mathbf{p},\mathbf{q})$ over $6$ elements. We assume
	the elements are indexed so that $p_i/q_i$ is monotonically decreasing.
	$\mathbf{p}$ and $\mathbf{q}$ are not homogeneous with respect to one
	another, and so the Lorenz curve bows out above the diagonal. \emph{(b)} An
	example of two pairs, $(\mathbf{p},\mathbf{q})$ and
	$(\mathbf{p}',\mathbf{q}')$, such that $(\mathbf{p},\mathbf{q})\succeq
	(\mathbf{p}',\mathbf{q}')$. Visually, this means that the second Lorenz
	curve is fully beneath the first and, therefore, closer to the line of
	homogeneity. \emph{(c)} Example where majorization does not hold. The
	extent of failure can be described by the \emph{dismajorization}, defined
	as the total $\mathbf{q}'$-probability associated with the
	majorization-breaching vertices. ``DM'' stands for dismajorization.
	}
	\label{fig:maj}
\end{figure*}

\subsection{Majorization and statistical mechanics}
\label{sec:maj}

To fully appreciate the results coming from the null models, we must compare
various distributions. To this end, we introduce a tool with a long tradition
that recently gained traction and found significant development in information
theory and statistical physics. The tool in question is
\emph{majorization}---more specifically, relative majorization
\cite{veinott_least_1971,marshall_inequalities_2011,Rene16a}. We describe the
necessary background below and also provide a primer in \cref{fig:maj}.

Given two probability distributions $\mathbf{p}=(p_i)$ and $\mathbf{q}=(q_i)$
defined over a finite set $\mathcal{S}$, we construct their {\em Lorenz curve}
$\ell_{\mathbf{p},\mathbf{q}}:[0,1]\rightarrow [0,1]$ as the piecewise convex
function connecting the points $(x_n,y_n)$:
\begin{align*}
  x_n = \sum_{m=1}^n p_{i_m} ~,
  \quad
  y_n = \sum_{m=1}^n q_{i_m}
  ~,
\end{align*}
where $(i_m)=(i_1,i_2,\dots)$ orders the set $\mathcal{S}$ so that
$p_{i_m}/q_{i_m}$ is monotonically decreasing in $m$. 

Given two pairs of distributions $(\mathbf{p},\mathbf{q})$ and
$(\mathbf{p}',\mathbf{q}')$, if $\ell_{\mathbf{p},\mathbf{q}}(x) \geq
\ell_{\mathbf{p}',\mathbf{q}'}(x)$ for all $x\in[0,1]$, then we say that
$(\mathbf{p},\mathbf{q})$ \emph{majorizes} $(\mathbf{p}',\mathbf{q}')$
\cite{veinott_least_1971}:
\begin{align*}
	(\mathbf{p},\mathbf{q})\succeq (\mathbf{p}',\mathbf{q}')
  ~.
\end{align*}

An intuitive application of majorization in fact arose in its first use as an
indicator of economic inequality \cite{lorenz_methods_1905}.  In this context,
if $\mathbf{p}$ describes the population distribution and $\mathbf{q}$ the
wealth distribution, the Lorenz curve completes statements of the type ``The
richest $x\%$ of the population holds $\ell_{\mathbf{p},\mathbf{q}}(x)\%$ of
the wealth.'' One country can be said to be definitively more unequal than
another if its Lorenz curve is always higher. Majorization generalizes this
relationship.

The connection between majorization and nonequilibrium statistical mechanics,
as well as its connection to our work, arises as a consequence of the
\emph{Blackwell-Sherman-Stein} (BSS) theorem
\cite{blackwell_comparison_1951,blackwell_equivalent_1953}:
$(\mathbf{p},\mathbf{q})\succeq (\mathbf{p}',\mathbf{q}')$ if and only if there
exists a stochastic matrix $\mathbf{T}$ such that
$\mathbf{T}\mathbf{p}=\mathbf{p}'$ and $\mathbf{T}\mathbf{q}=\mathbf{q}'$.
This connection is profound, given the frequent appearance of stochastic
matrices in statistical mechanics, information processing, stochastic
processes, game theory, and decision theory, and far more.

It has quite recently found significant application in the intersection between
information theory and nonequilbrium statistical mechanics. Actions taken upon
a thermodynamic system can be described as stochastic matrices over a system's
microstates. In this setting, it can be shown that any action (described by
stochastic matrix $\mathbf{t}$) that satisfies (i) energy conservation, (ii)
Liouville's theorem or information conservation, and (iii) access to a thermal
reservoir of temperature $T$ must obey the constraint:
\begin{align}
\label{eq:gibbs-stochasticity}
  \sum_j t_{i|j} \gamma_j(T)
  =\gamma_i(T)
  ~,
\end{align}
where $\gamma_i$ is the Boltzmann-Gibbs distribution:
\begin{align*}
  \gamma_i(T) = \frac{e^{- E_i/kT}}{Z(T)},\quad Z(T) = \sum_i e^{- E_i/kT}
\end{align*}
and $\mathbf{E}=(E_i)$ defines the energies of each microstate $i$
\cite{janzing_thermodynamic_2000}.

The significance of this observation is that even when operating on a
distribution $\mathbf{p}$ that is far from equilibrium---that is, \emph{not}
equal to $\boldsymbol{\gamma}(T)$---the actions we take must still satisfy
\cref{eq:gibbs-stochasticity}. Using the BSS theorem, we learn that a
distribution $\mathbf{p}$ can be physically transformed into another
$\mathbf{p}'$ using a bath at temperature $T$ only if
$(\mathbf{p},\boldsymbol{\gamma}(T))\succeq(\mathbf{p}',\boldsymbol{\gamma}(T))$.
It is said in this case that $\mathbf{p}$ \emph{thermo-majorizes} $\mathbf{p}'$
\cite{horodecki_fundamental_2013}. This connection between majorization and
thermodynamics has been used to derive a number of relations that leverage
thermodynamic fluctuations to extract work in the nanoscale, single-shot regime
\cite{brandao_resource_2013,horodecki_fundamental_2013,
brandao_second_2015,Rene16a,Lost18a}.

When majorization does not hold, it may hold approximately---a fact that can
still result in many of the same consequences of majorization. It will be
useful to have a quantification of dismajorization that allows us to
distinguish between small and large violations of majorization.

We define the \emph{dismajorization} $\mathrm{DM}\left[ (\mathbf{p},\mathbf{q});
(\mathbf{p}',\mathbf{q}') \right]$ of two pairs of curves in the following way.
Let $(i_m)$ be the same ordering of indices used above and let $x_n'$ and
$y_n'$ be defined in the same manner as $x_n$ and $y_n$ but for the pair
$(\mathbf{p}',\mathbf{q}')$. Finally, let $\mathcal{N}$ be the set of $n$ such that $y_{n}'>\ell_{\mathbf{p},\mathbf{q}}(x_n'))$. Then we define:
\begin{align*}
  \mathrm{DM}\left[
  (\mathbf{p},\mathbf{q});
  (\mathbf{p}',\mathbf{q}')
\right]
  = \sum_{n\in\mathcal{N}}q'_{i_n}
\end{align*}
as the total probability associated with the points where the second Lorenz
curve exceeds the first.

\section{Results}
\label{sec:results}

Leontief analysis of the null-model MRIO tables paired with the empirical labor
and $\CO_2$ distributions exposed a strong bias for high-intensity regions to
be net exporters and low-intensity regions to be net importers of embodied
$\CO_2$ emissions. This finding mirrors previous results
\cite{krausmann_global_2008, peters_production-based_2008, erb_embodied_2009,
moran_trading_2009, davis_consumption-based_2010, wiedmann_carbon_2010,
peters_growth_2011,davis_supply_2011, bergmann_bound_2013,
yu_tele-connecting_2013, alsamawi_employment_2014, simas_bad_2014,
dorninger_can_2015, liu_carbon_2015, bergmann_land_2016,
oita_substantial_2016}. The key difference is that our trade networks were
entirely randomized. Importantly, this suggests that our findings are not a
consequence of global trade relations, but an \emph{artifact resulting from
pairing unequal intensities with Leontief analysis}. Given that this is a
natural research strategy for the field, the conclusion is a cautionary lesson.

\Cref{sec:null} discusses the primary results from our null-model MRIO table.
\Cref{sec:ecomaj} then follows with a definition of \emph{eco-majorization}. We
show how, in the framework of Leontief analysis, it drives global flows of
trade. This leads us to hypothesize the conditions under which Leontief
analysis is biased towards eco-majorized results. \Cref{sec:relax} alters the
parameters of the null model to explore the consequences of relaxing these
conditions. This reveals a strong relationship between both
conditions---between eco-majorization and the directionality of embodied
$\CO_2$ flows.

\subsection{Flows in the null model}
\label{sec:null}
\begin{figure*}[t]
\centering
\includegraphics[width=1.0\textwidth]{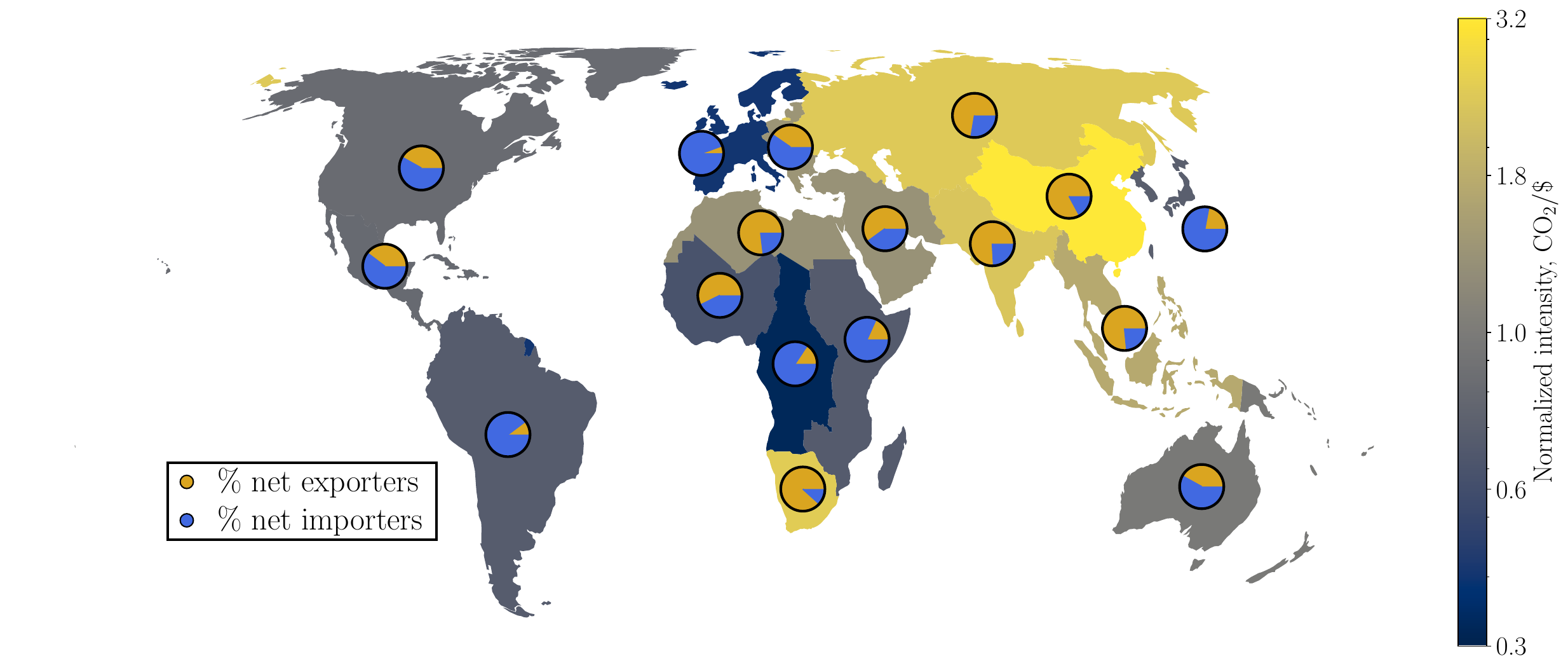}
\caption{Embodied $\CO_2$ flows: Two sets of information for each of 17
	megaregions. \emph{Regional color}: Normalized intensity
	$\hat{{f}}^{(\CO_2)}_r$ for each region $r$ with bright yellow indicating
	higher intensities and dark blue indicating lower. \emph{Regional pie
	charts}: Proportion of null models for which the given region was a net
	exporter (yellow) or a net importer (blue). A chart's amount of yellow
	corresponds to the quantity $\Xi^{(\alpha)}_r$---sample proportion of net
	positive exports; see text. This map uses the Equal Earth projection
	\cite{savric_equal_2019}.
	}
\label{fig:map}
\end{figure*}

We drew $1000$ samples from the null model with $\zeta_{C,i}$ and $\zeta_X$ set
at the baseline values $\bar{\zeta}_{C,i}$ and $\bar{\zeta}_X$. For each sample
from the null model, we juxtaposed the resulting attribution matrix with the
predetermined $\CO_2$ intensity $\hat{\mathbf{f}}^{(\CO_2)}$ and labor
intensity $\hat{\mathbf{f}}^{(\mathrm{L})}$ to calculate the embodied flows of
these impacts. For each region and impact, we calculated the proportion of
samples for which the net exports $\xi^{(\alpha)}_r$ were positive, denoting
the proportion $\Xi_r^{(\alpha)}$. In addition to being a function of the
region $r$, $\Xi_r^{(\alpha)}$ is dependent on the null model parameters as
well as the impact distribution $\hat{\mathbf{e}}^{(\alpha)}$, and may be
termed the \emph{null likelihood of net exports} for region $r$ with respect to
resource $\alpha$. \Cref{fig:map} displays the impact intensities in color and
the null export likelihood as a pie chart for each region with respect to
$\CO_2$.

Despite the null model having no preferred directionality between regions or,
for that matter, even any preferred tendency between imported and domestic
sources, one sees that high-intensity regions have a strong tendency to export
embodied $\CO_2$ while low-intensity regions have a strong tendency to import
embodied $\CO_2$. Moderately-intense regions do not exhibit bias.

The relationship between intensity $\hat{f}^{(\alpha)}_r$ and export likelihood
$\Xi_r^{(\alpha)}$ can be expressed using a nonlinear measure of correlation,
such as Kendall's $\tau$ \cite{kendall_1938}. We found that the correlation
between the two quantities for carbon was $\tau^{(\CO_2)}=0.68$. While, for
labor, the correlation was a remarkable $\tau^{(\mathrm{L})}=0.96$. Likely,
this is due to the fact that labor intensities are determined on the factor
level---consequently, there is no intra-regional variation in labor intensity
that confounds the relationship between intensity and global trade.

In this way, the null models demonstrate that complete randomization of social
factors in MRIO tables has little effect on the directionality of embodied
flows: they are still directed from high- to low-intensity regions. The
following section offers an explanation, via majorization, for how the
assumptions underlying Leontief analysis itself drive the correlation between
impact intensities and embodied flows.

\subsection{Eco-majorization and Leontief bias}
\label{sec:ecomaj}

Majorization's key benefit is that it readily explains the internal mechanics
of input-output analysis. In this way, the present use is yet another example
of generalizing thermodynamic logic to new settings. Such applications have,
for instance, already been powerfully applied to develop quantum resource
theories, which make frequent use of majorization to study entanglement and
other quantum properties as a nonfungible resources
\cite{Horo09a,janzing_thermodynamic_2000, brandao_resource_2013,
horodecki_fundamental_2013, brandao_second_2015}. 

The following, using it, shows that Leontief analyses tends to detect flows of
embodied impacts from high-intensity regions to low-intensity regions. The
effect is physically analogous to particles diffusing from high-density to
low-density regions. In this way, majorization connects these two settings.

Majorization is defined on probability vectors, whose total sum is normalized
to $1$, but we will for simplicity write unnormalized vectors in the
majorization pairs as a shorthand for the majorization of their normalized
forms. We will demonstrate that if the following conditions hold for an EE-MRIO
with impact $\alpha$:
\begin{enumerate}[label={(MRIO-\arabic*)},align=left]
\item \label{itm:mrio1} The regional impact intensities
	$\hat{\mathbf{f}}^{(\alpha)}$ are highly heterogeneous and
\item \label{itm:mrio2} The regional deficit
	$\mathbf{\hat{x}}-\mathbf{\hat{y}}$ is small as a proportion of 
	regional income across regions,
\end{enumerate}
then with high probability Leontief analysis results in (or approximately
results in) the majorization:
\begin{align}
\label{eq:eco-maj}
	\left(\mathbf{\hat{e}^{(\alpha)}},
	\mathbf{\hat{y}}\right)
	\succeq \left(\mathbf{\hat{a}^{(\alpha)}},
	\mathbf{\hat{y}}\right)
	~.
\end{align}
As this phenomenon links both ecological impacts and economic activity levels,
we call this relation \emph{eco-majorization}, where the prefix may refer to
either. We note that both assumptions hold for the GTAP 8 dataset.

Since the regional income distribution $\mathbf{\hat{y}}$ plays a role
similar to the thermodynamic Gibbs distribution, the stated relation tells us
that the embodied impacts $\hat{\mathbf{a}}^{(\alpha)}$ are more similarly
distributed to the regional incomes than to the local impacts
$\hat{\mathbf{e}}^{(\alpha)}$. This necessitates transferring embodied impacts
from high-intensity regions, where distributional imbalance is most positive,
to low-intensity regions, where it is most negative.

\begin{figure}[t]
\centering
\includegraphics[width=0.5\columnwidth]{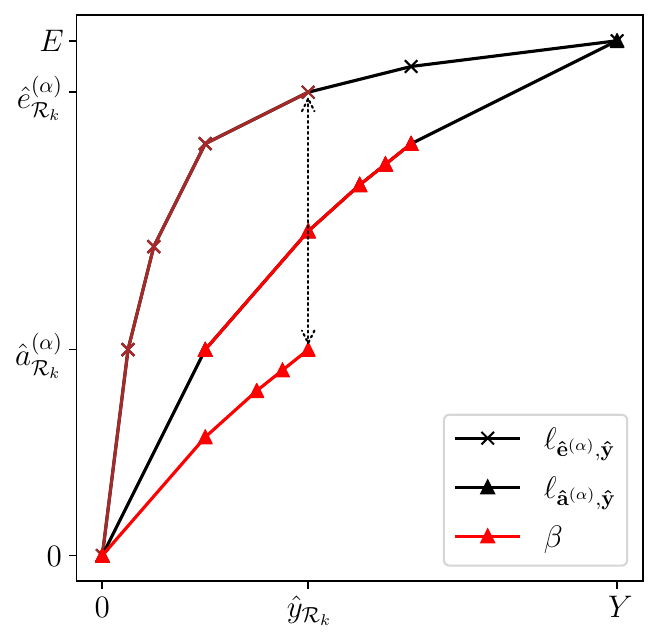}
\caption{Most intense regions must be net exporters: Graphical proof that, as
	long as eco-majorization holds, the $k$ most intense regions
	$\mathcal{R}_k$ will have $\mathbf{\hat{a}}^{(\alpha)}_{\mathcal{R}_k}\leq
	\mathbf{\hat{e}}^{(\alpha)}_{\mathcal{R}_k}$. This rests on the fact that
	the partial Lorenz curve $\beta$ corresponding only to the regions
	$\mathcal{R}_k$ must be below the Lorenz curve
	$\ell_{\mathbf{\hat{a}}^{(\alpha)},\hat{\mathbf{y}}}$. And that, in turn,
	falls below $\ell_{\mathbf{\hat{e}}^{(\alpha)},\hat{\mathbf{y}}}$. 
	}
\label{fig:picproof}
\end{figure}

We quantify the previous statement using net exports
$\boldsymbol{\xi}^{(\alpha)}$ and impact ratios
$\boldsymbol{\rho}^{(\alpha)}_r$. Let $\mathcal{R}_k$ be that subset of regions
containing the $k$ regions with the highest values of $\hat{f}^{(\alpha)}_r$.
Then it can be shown that, if eco-majorization holds:
\begin{align}
\label{eq:flows-maj}
  \sum_{r\in \mathcal{R}_k} \xi^{(\alpha)}_r  \geq 0 ~ ~\text{and} ~ ~
  \sum_{r\in \mathcal{R}_k} \frac{\hat{e}^{(\alpha)}_r}{E}\rho^{(\alpha)}_r
  \leq 1
  ~.
\end{align}
Both indicators then tell us that regions in $\mathcal{R}_k$ must be net
exporting or, at least, never net importing. \Cref{fig:picproof} presents a
visual proof using Lorenz curves. Since a Lorenz curve is monotonically
decreasing in slope, the curve $\beta(x)$ formed by taking only a subset of
segments must always be lower in height than the original curve. From this it
can be seen that when $\left(\mathbf{\hat{e}^{(\alpha)}},
\mathbf{\hat{y}}\right) \succeq \left(\mathbf{\hat{a}^{(\alpha)}},
\mathbf{\hat{y}}\right)$, we must have $\sum_{r\in \mathcal{R}_k}
\mathbf{\hat{a}}_r\leq \sum_{r\in \mathcal{R}_k}\mathbf{\hat{e}}_r$, from which
Eqs. \eqref{eq:flows-maj} hold.

Eco-majorization, then, places rigid constraints on the directionality of
trade flows. Our null model simulation decisively demonstrates it is at play in
the observed relationship between intensity and exports: embodied labor flows
were eco-majorized for $100\%$ of the simulated networks and embodied $\CO_2$
flows were eco-majorized for $72\%$ of the simulated networks.

These results suggest that while eco-majorization is not guaranteed, it is a
highly ubiquitous phenomenon among randomly generated MRIO tables.  The issue,
we argue, rests in the presence of the two conditions \ref{itm:mrio1} and
\ref{itm:mrio2}. 

Due to \cref{eq:attr-income,eq:attr-impact}, the BSS theorem automatically
entails:
\begin{align}
\label{eq:full-maj}
\left(\mathbf{{e}^{(\alpha)}}, \mathbf{{v}}\right)
	\succeq \left(\mathbf{\hat{a}^{(\alpha)}},
	\mathbf{\hat{x}}\right)
  ~.
\end{align}
This differs from \cref{eq:eco-maj} in two key respects. First, the lefthand
side refers to the sectoral impact distribution $\mathbf{{e}^{(\alpha)}}$ and
the factor income distribution $\mathbf{{v}}$ rather than the regionalized
distributions $\mathbf{\hat{y}^{(\alpha)}}$ and $\mathbf{\hat{y}}$,
respectively. We call this difference \emph{regionalization}. Second, the
righthand side uses the spending distribution $\mathbf{\hat{x}}$ instead of the
income distribution $\mathbf{\hat{y}}$. This is of little concern issue when
the regional deficit is small, as assumed in \ref{itm:mrio2}. By the nature of
Lorenz curves, small changes in the underlying distributions result in
correspondingly small changes in curve's shape.

The subtlety, and the only reason why $100\%$ of simulated networks do not
display majorization for all impacts, arises from regionalization: 
\cref{eq:full-maj} shows that majorization
holds for the full sectoral distributions, but does
not say anything about regional distributions. It is entirely \emph{possible}
that \cref{eq:full-maj} may be true and \cref{eq:eco-maj} may be false.
The crux of this issue is, in fact, the same as a major point of
thermo-majorization theory: namely, that when the majorizing pair of
distributions are coarse-grained, majorization might no longer hold. Notably,
many recent results on the work cost of driving systems away from thermodynamic
equilibrium exploit this phenomenon \cite{brandao_second_2015,Rene16a}.

We are not interested here in how to induce this phenomenon. Rather, we are
interested in why it does not appear to naturally arise in either the existing
trade data or the null model. Our argument is that condition \ref{itm:mrio1}
significantly constrains the possible configurations that may result in
a violation of \cref{eq:eco-maj}.

The argument is as follows. 
For a set $\mathcal{S}$ of regions, define:
\begin{align*}
  \hat{y}_{\mathcal{S}} := \sum_{s\in\mathcal{S}} \hat{y}_s ~,\quad
  \hat{f}^{(\alpha)}_\mathcal{S} := 
  \frac{\sum_{s\in\mathcal{S}}\hat{e}^{(\alpha)}_s/E}
  {\hat{y}_{\mathcal{S}}/Y} ~,~\text{and} ~ ~
  \hat{f}^{(\alpha)'}_\mathcal{S} := 
  \frac{\sum_{s\in\mathcal{S}}\hat{a}^{(\alpha)}_s/E}
  {\hat{y}_{\mathcal{S}}/Y}
  ~.
\end{align*}
Further, let $\hat{y}_{k}:=\hat{y}_{\mathcal{R}_k}$
for simplicity. To violate majorization
there must be a set $\mathcal{S}$ of regions such that:
\begin{align}
\label{eq:no-maj}
  \hat{f}^{(\alpha)'}_\mathcal{S} 
  \geq \left(\frac{\hat{y}_{\mathcal{S}}-
  \hat{y}_k}{\hat{y}_{k+1}-\hat{y}_k}\right)
  \hat{f}^{(\alpha)}_{\mathcal{R}_k}
  +\left(\frac{\hat{y}_{k+1}-\hat{y}_{\mathcal{S}}
  }{\hat{y}_{k+1}-\hat{y}_k}\right)
  \hat{f}^{(\alpha)}_{\mathcal{R}_{k+1}}
  ~,
\end{align}
where $k$ is the unique integer such that $\hat{y}_k\leq \hat{y}_{\mathcal{S}}<
\hat{y}_{k+1}$. (This simply restates the definition of majorization via Lorenz
curves.) We can actually suppose without loss of generality that
$\hat{y}_k=\hat{y}_{\mathcal{S}}$. This can be achieved by splitting regions
into smaller but structurally identical subregions. So, \cref{eq:no-maj} can be
expressed more simply as $\hat{f}^{(\alpha)'}_\mathcal{S} \geq
\hat{f}^{(\alpha)}_{\mathcal{R}_k}$.

Now, we may rewrite $\hat{f}^{(\alpha)'}_\mathcal{S}$ as:
\begin{align}
\label{eq:slope}
  \hat{f}^{(\alpha)'}_\mathcal{S}=
  \sum_{\substack{r\in\mathcal{R}\\i\in\mathcal{I}_0}}
  \frac{\hat{A}_{ri,s}{y}_{ri}}{\hat{y}_{\mathcal{S}}}
  f^{(\alpha)}_{ri}
  ~.
\end{align}
This considerably constrains the structure of matrices $\hat{\mathbf{A}}$ that
yield a large value for $\hat{f}^{(\alpha)'}_\mathcal{S}$. Specifically, if
$\hat{f}^{(\alpha)'}_\mathcal{S} \geq \hat{f}^{(\alpha)}_{\mathcal{R}_k}$, then
$\hat{A}_{ri,s}$ either must put great weight on the most intense
sectors within the regions of $\mathcal{R}_k$ or it must draw from similarly
intense sectors that may, with small probability, have arisen in less intense
regions. In either case, $\hat{A}_{ri,s}$ must give high weight 
to sectors $(r,i)$ with
intensities
$f_{ri}^{(\alpha)}$ that exceed the average intensity of the highest $k$
regions: $f_{ri}>\hat{f}^{(\alpha)}_{\mathcal{R}_k}$.

The Markov inequality states that for any 
positive random variable $X$ with mean value 
$\hat{x}$, the probability that an instance exceeds the mean
by a proportion $\beta$
is constrained \cite{Cove06a}:
\begin{align*}
  \Prob{(X\geq \beta \hat{x})} \leq \frac{1}{\beta}
  ~.
\end{align*}
Then for each region $r$, the weight (under $\mathbf{v}$) that a given
sectoral intensity exceeds $\hat{f}_k^{(\alpha)}$ by a proportion $\beta$ is:
\begin{align*}
  \sum_{i: f_{ri}>\beta\hat{f}_{\mathcal{R}_k}^{(\alpha)}}
  v_{ri}
  \leq \frac{q_r \hat{f}^{(\alpha)}_r}
  {\beta\hat{f}_{\mathcal{R}_k}^{(\alpha)}}
  ~.
\end{align*}
Thus, due to the Markov inequality, high-intensity sectors are
suppressed in weight, in a manner determined by the relative proportions
of intensities between regions. When condition \ref{itm:mrio2} holds---that
is, the regional intensities are highly heterogeneous---this
suppression is strengthened.

To counter this suppression in \cref{eq:slope}, $\hat{A}_{ri,s}$ must place
extremely high relative weight on high-intensity sectors. In this case,
however, very little weight remains to distribute among other sectors.
Combinatorially, then, matrices $\hat{\mathbf{A}}$ violating regional
majorization occupy a relatively small niche in the space of all
configurations.

To summarize, as a consequence of the BSS theorem and fundamental facts of
Leontief analysis, \cref{eq:full-maj} must hold for any EE-MRIO table. When
assumptions \ref{itm:mrio1} and \ref{itm:mrio2} hold, implying heterogeneity of
regional intensities and small regional deficits, \cref{eq:full-maj} further
supports eco-majorization \cref{eq:eco-maj} by constraining the possible
configurations which are not eco-majorized.

In the analogous thermodynamic setting, the experimenter (a Maxwell's
``demon'') may intentionally configure matrices that violate majorization after
coarse-graining. In the setting of global trade, however, such a matrix must
come about as the result of a strict bias among some regions to only consume
high-intensity products. This is hardly realistic: Even at a national level,
imports are a function of the variegated needs of multiple consumers and
corporations. And, they necessarily draw their consumption from high-intensity
and low-intensity industries. Regions, in short, do not operate as Maxwellian
demons---at least, not in regards to their bulk imports. 

Furthermore, as the null model is symmetrically generated without knowledge of
local intensities, the null model is not be likely to draw models from the
small niche required significantly violate majorization. Indeed, it is worth
noting that our null model effectively acts as a Monte-Carlo model, calculating
the total probability mass of the configuration space where majorization is
violated.

\begin{figure*}[t]
\centering
\includegraphics[width=0.7\textwidth]{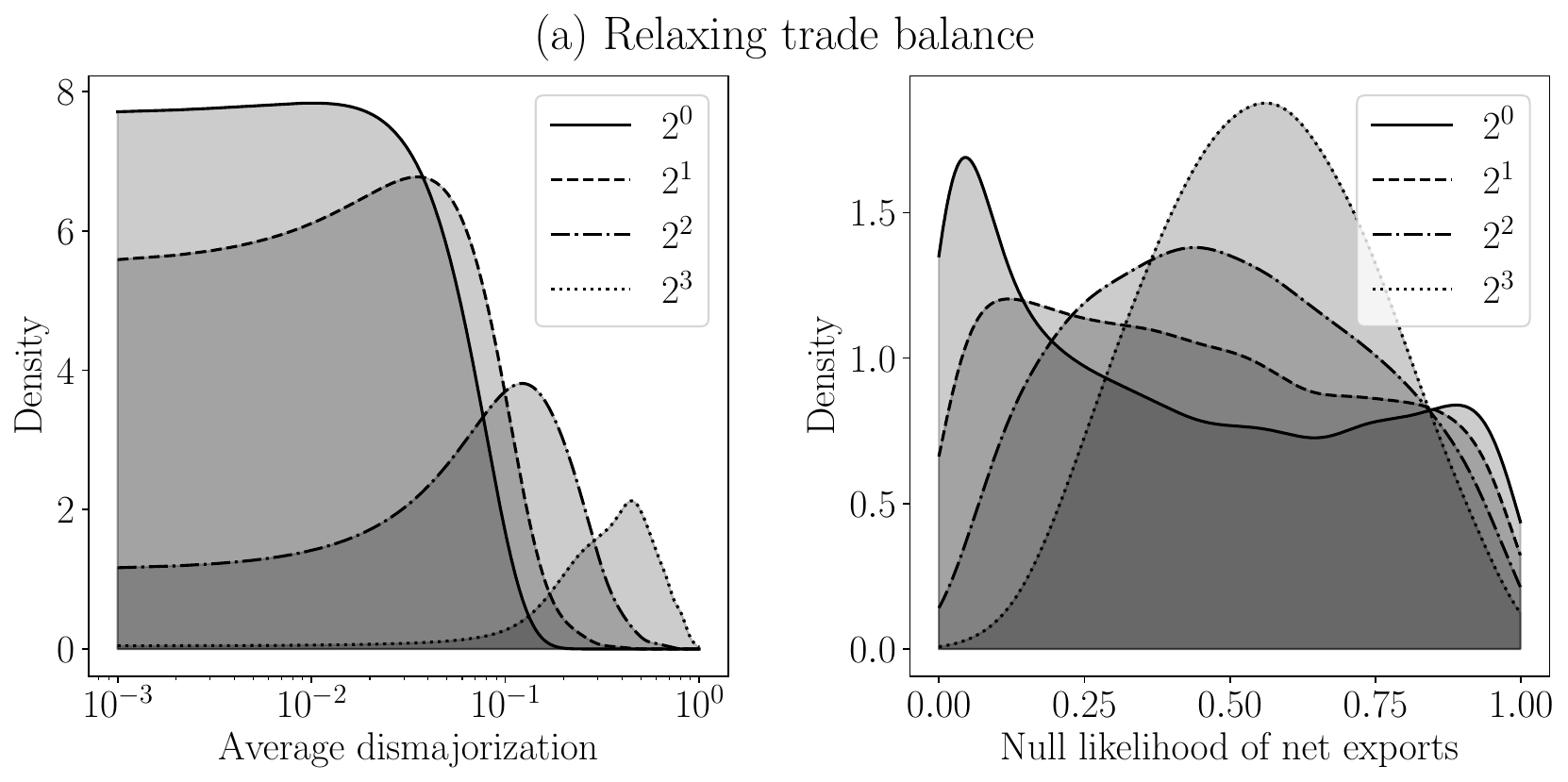}
\includegraphics[width=0.7\textwidth]{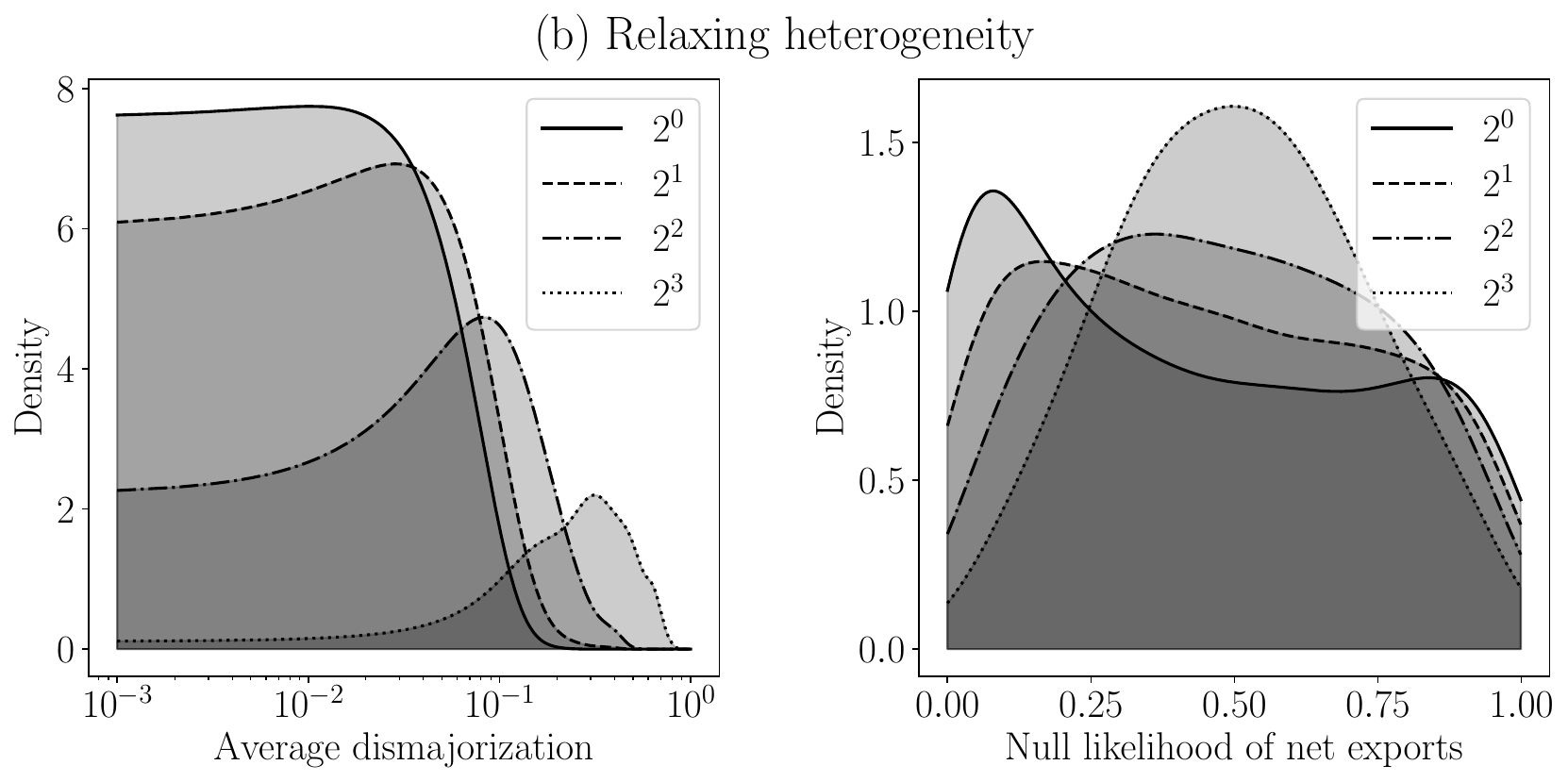}
\caption{Emergence of eco-majorization: The results of varying the conditions
	that, when paired with Leontief analysis, result in eco-majorization.
	\emph{(a)} Relaxing trade balance: Four suites of ensembles where the trade
	balance parameter was taken as $\zeta_X=\sigma^{-1}\zeta_X$,
	$\sigma=2^0,2^1,2^2,2^3$. As $\sigma$ grows, so does the ensemble's overall
	trade deficit. A Gaussian kernel density was computed for the dismajorization
  and the null likelihood of net export
  over the null model samples.
  The left plot demonstrates that trade deficit's increase has the effect of
	increasing the average dismajorization. The right plot demonstrates
  that this also has the impact of shifting the net export density
  from being bimodal to unimodal.
  \emph{(b)} Relaxing heterogeneity:
	Four suites of ensembles where the impact heterogeneity parameter was taken
	as $\zeta_U=\sigma\zeta_U$, $\sigma=2^0,2^1,2^2,2^3$. As $\sigma$ grows,
	heterogeneity decreases. This has similar effects on the dismajorization
  density plot and the net export density plot.
	}
\label{fig:graphs}
\end{figure*}

\subsection{Relaxing assumptions}
\label{sec:relax}

To verify that conditions \ref{itm:mrio1} and \ref{itm:mrio2} are indeed
responsible for the appearance of majorization and, consequently, the coupling
between intensities and embodied flows, we made two modifications to the null
model. First, we introduced null-impact intensities. Second, we varied
parameters of both the MRIO null model and the impact intensity null model to determine
their effects. We performed these in two separate trials.

For the first, we generated $1000$ unobtainium intensity functions with
parameter $\zeta_U=\bar{\zeta}_U$. For each scaling factor
$\sigma=1,2,4,8$, we generated $250$ MRIO tables from the null model with
$\boldsymbol{\zeta_C}=\boldsymbol{\bar{\zeta}_C}$ and $\zeta_X = \sigma^{-1}
\bar{\zeta}_X$. This results in $4$ ensembles of $250,000$ EE-MRIO tables each,
with increasing trade imbalances from one ensemble to the next.

For the second, we flipped the structure of this approach, generating $1000$
MRIO tables from the null model with
$\boldsymbol{\zeta_C}=\boldsymbol{\bar{\zeta}_C}$ and $\zeta_X =
\bar{\zeta}_X$. Then, for each scaling factor $\sigma=1,2,4,8$, we generated
$250$ unobtainium intensities with parameter $\zeta_U=\sigma \bar{\zeta}_U$.
This results in the same number of ensembles, but with decreasing heterogeneity
of intensities.

To quantify the effect of the changes in parameter on majorization, we used
dismajorization 
$\mathrm{DM}\left[
  (\mathbf{\hat{e}}^{(\mathrm{U})},\mathbf{\hat{y}});
  (\mathbf{\hat{a}}^{(\mathrm{U})},\mathbf{\hat{y}})
\right]$,
as defined in \cref{sec:maj}. For the first suite of ensembles, for every
sampled MRIO table we calculated the average dismajorization over all
unobtainum distributions. Then, a density plot of average dismajorizations over
all samples with a given scaling factor $\sigma$ was computed. The resulting
density functions are shown in \cref{fig:graphs}. Similarly, for the second
suite, for every sampled unobtainium distribution we calculated the average
dismajorization over all MRIO tables. Density plots were similarly taken for
each scaling factor $\sigma$. The resulting density functions are also shown in
\cref{fig:graphs}. We find in each case that the likely dismajorization
increases dramatically as the assumptions \ref{itm:mrio1} and \ref{itm:mrio2}
are relaxed.

For both suites, we also sought to examine the impact of the scaling factors on
trade-flow directionality. To this end, we calculated the null likelihoods of
net exports $\Xi_r$ for each region and each unobtainium distribution
$\hat{\mathbf{e}}^{(U)}$. We made a density plot of observed values of $\Xi_r$
for each value of $\sigma=1,2,4,8$. At the baseline parameter values
($\sigma=1$), this density plot is bimodal, with one mode close to zero
(countries that tend to be net importers) and one mode close to one (those that
tend to be net exporters). However, as the baseline values are altered to relax
assumptions, the density of $\Xi_r$ becomes unimodal---the sharp distinction
between exporters and importers vanishes. This is true as either assumption is
relaxed.

These results complement those of the previous section. In \Cref{sec:ecomaj} we
showed that that the assumptions \ref{itm:mrio1} and \ref{itm:mrio2} are
together sufficient for majorization to be predominant; Fig. \ref{fig:graphs}
shows that they are each necessary.  Additionally, relaxing either assumption
(and the consequent irrelevancy of majorization) diminishes an initially strong
dichotomy between exporting and importing nations, as the density of $\Xi_r$
goes from bimodal to unimodal.

\section{Discussion \& Concluding Remarks}
\label{sec:discussion}

The two assumptions of Leontief analysis, \ref{itm:L1} and \ref{itm:L2}, are
critically important for understanding (i) the emergence of majorization in
this setting and (ii) how exploring the validity of Leontiefian assumptions has
great merit in assessing its results' usefulness. The ``foothold'' that
majorization makes in our analysis begins with
\cref{eq:attr-income,eq:attr-def}:
\begin{align*}
  \mathbf{v}\hat{\mathbf{A}} = \hat{\mathbf{x}} ~ ~\text{and} ~ ~
  \mathbf{e}^{(\alpha)}\hat{\mathbf{A}} = \hat{\mathbf{a}}^{(\alpha)}
  ~.
\end{align*}
These equations relate factor incomes to the regional spending and production impacts to attributed impacts. Since both use the same attribution matrix $\hat{\mathbf{A}}$, the initial majorization relation \cref{eq:full-maj} holds. Our mathematical analysis rests on this fact.

In information theory, stochastic matrices act as lossy channels that transmit
information in a degraded condition. This leads distinct channel inputs to
become more similar. While we sorted out the subtleties here, it is primarily
for this reason that embodied emissions become more similar to global income:
they are both channeled through the same lines of flow, determined by the
single matrix $\mathbf{\hat{A}}$.

However, employing the same attribution matrix---whose primary determinant is
\emph{monetary} flows---to also drive embodied impact flows is a possibility
only allowed by \ref{itm:L1} and \ref{itm:L2}. The homogeneity of products and
prices allows us to assume that monetary flows are entirely sufficient to
reconstruct the commodity chains in which embodied flows are materialized.
Without these assumptions, one would have a unique attribution matrix
$\hat{\mathbf{A}}^{(\alpha)}$ for each impact $\alpha$, distinct from the
monetary attribution matrix. Majorization would no longer necessarily hold. And
so, it would no longer drive the relationship between local intensities and
embodied flows. In this way, our results suggest that the outcomes of Leontief
analysis are highly dependent on the assumptions made. Rather than a neutral
tool of analysis, Leontiefian methods embody significant ideological
consequences.

Consider in this light the carbon leakage hypothesis. Essentially, firms from
high-income countries foist the direct carbon costs of their production (which
feeds local consumption) onto lower-income countries. This maintains
consumption patterns while lowering compliance costs by superficially adhering
to climate treaties to which they are signatories
\cite{peters_growth_2011,davis_supply_2011}. 

When this extends beyond carbon to other environmental impacts, the flow of
embodied impacts from low- to high-income countries is \emph{ecologically
unequal exchange}---a major topic in modern geography and ecological economics
\cite{hornborg_unequal_2003, rice_ecological_2007, smith_trade_2012,
jorgenson_sociology_2012-1}. One can even consider impacts such as labor-time.
This leads to the more traditional hypothesis of unequal exchange
\cite{emmanuel_unequal_1972} of exploited labor from low- to high-income
countries. Each of these hypotheses rests on fundamental assumptions about the
social and economic power relations between nations and regions.

Multiregional input-output tables and Leontief analysis have been put to use
evaluating these hypotheses previously, frequently in tandem with
quantities such as the net exports $\boldsymbol{\xi}^{(\alpha)}$ and
consumption-to-production ratio $\boldsymbol{\rho}^{(\alpha)}$ which we have
analyzed here
\cite{erb_embodied_2009,
davis_consumption-based_2010, bergmann_bound_2013, moran_does_2013,
yu_tele-connecting_2013, alsamawi_employment_2014, simas_bad_2014,
dorninger_can_2015, liu_carbon_2015, bergmann_land_2016, oita_substantial_2016}. 
Our results directly bear on these previous studies. We showed that
these quantities are, in fact, strongly driven by the assumptions of Leontief
analysis and are largely independent of the relational data within the MRIO
tables used. This calls into question the validity of Leontief analysis as a
tool for empirically verifying hypotheses of unequal exchange,
contributing a new perspective to previous critiques of this application
\cite{dorninger_can_2015}.

Employing MRIO tables and Leontief analysis for empirical purposes must be done
with caution. To this end, we recommend an approach based on that taken
here. Specifically, for this we make two contributions. First, use modern tools
from information theory and statistical physics to better understand the
consequences of methods like Leontief analysis, embedded as they are with
numerous stochastic matrices and distributional relationships. Second,
frequently consult with null models. This will aid in disentangling data
structures, mathematical artifacts, and hypotheses, as otherwise these can be
quite difficult to tease apart when using sophisticated modeling assumptions.
This recommendation, of course, extends beyond MRIO tables and Leontief
analysis. However, applying this approach to other studies of carbon
accounting, environmental impacts, and ecologically unequal exchange will
remain for future work.

\section*{Acknowledgments}
\label{sec:acknowledgments}

The authors thank Luke Bergmann and Fushing Hsieh for helpful discussions and
the Telluride Science Research Center for hospitality during visits and the
participants of the Information Engines Workshops there. As a faculty member,
JPC similarly thanks the Santa Fe Institute. This material is based upon work
supported by, or in part by the U.S. Army Research Laboratory and the U. S.
Army Research Office under grants W911NF-18-1-0028 and W911NF-21-1-0048 and by
the UC Davis CeDAR Innovative Data Science Seed Funding Program.

\section*{Competing interests statement}

The authors have no competing interests to declare.

\section*{Reproducibility statement}

To contribute to the methods' reproducibility, we provide the scripts and
Jupyter Notebooks used to perform our analysis in a Github repository
\cite{Github-NETACAM}. The code makes use of the package {\tt stoclust}, a tool
for ensemble-based statistical analysis \cite{Github-stoclust}. The source data
must be acquired from GTAP 8, the most recent freely accessible version of GTAP
\cite{GTAP,narayanan_g_global_2012}.

\appendix
\onecolumngrid
\clearpage
\begin{center}
\large{Supplementary Materials}\\
\vspace{0.1in}
\emph{\ourTitle}\\
\vspace{0.1in}
{\small
Samuel P. Loomis, Mark Cooper, and James P. Crutchfield
}
\end{center}

\setcounter{equation}{0}
\setcounter{figure}{0}
\setcounter{table}{0}
\setcounter{page}{1}
\makeatletter
\renewcommand{\theequation}{S\arabic{equation}}
\renewcommand{\thefigure}{S\arabic{figure}}
\renewcommand{\thetable}{S\arabic{table}}

In these appendices we discuss further background on input-output analysis
and the construction of our null model.

\section{Input-output analysis, in further detail}
\label{app:io-analysis}
\subsection{Basic input-output analysis}
An \emph{economy} $(\mathcal{I},\mathcal{V},\mathcal{D})$ is composed of finite
sets of industrial sectors $\mathcal{I}$, value-added sectors $\mathcal{V}$,
and final demand sectors $\mathcal{D}$. Value-added sectors usually include
factors of production, such as labor, capital, land, and natural resources.
Final demand sectors indicate the various forms of consumption: traditionally,
private consumption, government spending, and business investment.

An economy's operation is cast as various kinds of \emph{flow} between the
sectors. To capture these for an economy, an input-output table is defined as
the triple $(\mathbf{Z},\mathbf{V},\mathbf{D})$ of matrices with interindustry
flows $Z_{ij}$ $(i,j\in\mathcal{I})$; value-added flows
$V_{ui}$ ($i\in\mathcal{I}$, $u\in\mathcal{V}$); and final-demand
flows $D_{ia}$, ($i\in\mathcal{I}$, $a\in\mathcal{D}$).
Interindustry flows describe transactions between industrial firms; value-added
flows describe factor returns such as wages, profits, and rent; final-demand
flows describe the direct spending by individuals, governments, and businesses
on consumable commodities, services, and fixed capital.

Each matrix component describes the flow of money from the column sector to the
row sector over a given time period (typically a year). For instance, $Z_{ij}$
describes the total flow of money from sector $j$ to sector $i$. Each
industrial sector is assumed to be balanced, so that the total outlays equal
the total output:
\begin{align}
\label{eq:balance-app}
  \underbrace{
    \sum_{u\in\mathcal{V}} V_{ui} 
  + \sum_{j\in\mathcal{I}} Z_{ji}
  }_{
    \text{Outlays}
  }
  = \underbrace{
    \sum_{j\in\mathcal{I}} Z_{ij}
    + \sum_{a\in\mathcal{D}} D_{ia}
  }_{
    \text{Outputs}
  }
\end{align}
We define, respectively, the total output 
$z_i$ of industry $i$, total income
 total demand $v_i$ of industry $i$, 
 and total value-added $d_i$ by industry $i$ as follows:
\begin{align}
\label{eq:defs-app}
  z_i := \sum_{j\in\mathcal{I}} Z_{ij} + \sum_{a\in\mathcal{D}} D_{ia},
  v_i := \sum_{u\in\mathcal{V}} V_{ui},\text{~and~}
  d_i := \sum_{a\in\mathcal{D}} D_{ia}
  ~.
\end{align}
We additionally define the total income as $Y:=\sum_i v_i$,
which is necessarily equal to the total spending $\sum_i d_i$
by \cref{eq:balance-app}.

A primary use of input-output analysis is to attribute the impacts of various
activities to each of the final demand sectors. This provides a useful way to
conceptualize the complex economy's interconnected causal relationships. To do
this, we first define the technical coefficients $C_{ij}$ as
$C_{ij} := Z_{ij}/z_j$.
These specify the outlays on activity $i$ required to produce a single monetary
unit of output in sector $j$. \Cref{eq:balance-app}'s balance condition can
then be written in matrix form as
$\mathbf{z} = \mathbf{C}\mathbf{z} + \mathbf{d}$,
a linear algebra problem whose solution (for $\mathbf{z}$) is:
\begin{align}
\label{eq:leontief-solution-app}
  \mathbf{z} = \left(\mathbf{I}-\mathbf{C}\right)^{-1} \mathbf{d}
  ~,
\end{align}
where $\mathbf{I}$ is the identity matrix and one uses the matrix inverse.
Written more explicitly we have:
\begin{align}
\label{eq:leontief-expanded-app}
  z_i = \sum_{a\in\mathcal{D}} \left[\left(\mathbf{I}-\mathbf{C}\right)^{-1} \mathbf{D}\right]_{ia}
  ~.
\end{align}
This expresses the total output as a column-sum of the matrix
$\left(\mathbf{I}-\mathbf{C}\right)^{-1} \mathbf{D}$. The sum allows us to
break sector $i$'s total output into parts, each attributed to a particular
final demand $a\in\mathcal{D}$. The attribution matrix $\mathbf{A}$:
\begin{align}
\label{eq:attribution-app}
  A_{ia} := \frac{\left[\left(\mathbf{I}-\mathbf{C}\right)^{-1} \mathbf{D}\right]_{ia}}{z_i}
  ~,
\end{align}
describes, for each dollar of output in sector $i$, how much of that dollar is
attributed to final demand $a$. Determining these attributions is called {\em
Leontief analysis} after its originator \cite{leontief_economy_1991,leontief_essays_1966}.\footnote{
  The analysis may be reversed to attribute outputs to factors $\mathcal{Y}$
  rather than to final demands $\mathcal{D}$. While not explicitly considered
  here, the results derived for demand-based accounting apply symmetrically to
  factor-based accounting.
}

It will be important, later, to note that $\sum_a A_{ia} = 1$. When this
property holds for a matrix, in addition to the condition of nonnegative
components, we say the matrix is \emph{stochastic}. It has considerable
importance for majorization.

The utility of Leontief analysis rests on two main assumptions
\cite{schaffartzik_environmentally_2014}:
\begin{enumerate}[label={(L-\arabic*)},align=left]
\item \label{itm:L1-app} \textbf{Sectors produce homogeneous products}: Due to this, we
	do not reweight the technical coefficients to reflect differences between
	the purchasing sectors---they purchase the same item.
\item \label{itm:L2-app} \textbf{Sectoral products are homogeneously priced}: Every buyer
	pays the same unit price. This again allows using the technical
	coefficients without modification to reflect differences in the inputs
	required per dollar for different purchasers. 
\end{enumerate}
The text discusses these further.

\subsection{Multiregional models}
\label{sec:multireg}

Multiregional input-output (MRIO) tables deepen the structure of input-output
tables by dividing sectors into regions \cite{leontief_essays_1966}.
Specifically, we suppose there is a finite set $\mathcal{R}$ of regions and
each set of sectors is organized as $\mathcal{I} = \mathcal{R}\times
\mathcal{I}_0$, $\mathcal{V} = \mathcal{R}\times \mathcal{V}_0$, and
$\mathcal{D} = \mathcal{R}\times \mathcal{D}_0$, where $\mathcal{I}_0$,
$\mathcal{V}_0$, and $\mathcal{D}_0$ are the regional-level industrial,
value-added, and final-demand sectors, respectively. An economy with this
structure is said to be a \emph{multiregional economy}.

The matrices and vectors described above can be adapted to this geographic
picture by replacing each individual index $i\in\mathcal{I}$ (or others) with
the pair $(r,i)$, $r\in\mathcal{R}$ and $i\in\mathcal{I}_0$.  This corresponds
to re-envisioning the matrices and vectors as block-matrices and block-vectors,
with rows and columns organized by regional blocks. See \cref{fig:io} in the
main text.

Simplifications are often imposed on MRIO tables. For instance, the
inter-industry flows $Z_{ri,sj}$ may involve considerable inter-regional
interaction, but it is typically supposed that $V_{ru,si} = 0$ and
$D_{ri,sa}=0$ whenever $r\neq s$. In other words, regional factors are paid
directly by a same-region industrial sector and regional consumption purchases
directly from a same-region sector. (The direct consumption of imports is
addressed by introducing intra-regional importing sectors to mediate the
inter-regional interaction, usually doubling
the size of $\mathcal{I}_0$.) This makes $\mathbf{V}$ and $\mathbf{D}$
block-diagonal.

Three concepts are important when identifying majorization.
First, while global income and global spending equal one another, it is not
necessarily the case that regional income and regional spending are equal. In
fact, this difference is directly related to the trade deficit, by
\cref{eq:balance-app}. Denoting the regional income $\hat{y}_r := \sum_i v_{ri}$, 
the regional spending $\hat{x}_r := \sum_i d_{ri}$, 
and the inter-regional trade
$\hat{Z}_{rs} := \sum_{ij} Z_{ri,sj}$, we have:
\begin{align*}
  \underbrace{
    \hat{y}_r - \hat{x}_r
  }_{\text{Income -- Spending}}
  = \underbrace{
    \sum_{s\in \mathcal{R}} \left(\hat{Z}_{rs}-\hat{Z}_{sr}\right)
  }_{\text{Exports -- Imports}}
  ~.
\end{align*}
We will refer to $\mathbf{\hat{y}}-\mathbf{\hat{x}}$ as simply the
\emph{regional deficit vector}. Its properties
will be important in our study of majorization
in Leontief analysis.

Second, when using MRIO tables to attribute economic activities to their corresponding
demands, considerably more focus is given to the region of demand than the
actual sector. Presently, this is our entire concern. We therefore define the
regional attribution matrix $\mathbf{\hat{A}}$ as:
\begin{align*}
  \hat{A}_{ri,s} := \sum_{a\in\mathcal{D}_0} A_{ri,sa}
  ~,
\end{align*}
where $\mathbf{A}$ is the block-matrix form of the attribution matrix defined
in \cref{eq:attribution-app}. In $\mathbf{\hat{A}}$, the rows are block-structured
by regions, while the each column directly corresponds to a unique region.
$\mathbf{\hat{A}}$ retains the stochastic property.

Third, a profoundly important identity emerges when applying the regional
attributions matrix to the value-added vector from \cref{eq:defs-app}: we arrive at
the regional spending vector $\hat{x}_r$:
\begin{align}
  \mathbf{v}\mathbf{\hat{A}} = \mathbf{\hat{x}}
  ~.
\label{eq:attr-income-app}
\end{align}
This follows from the relation:
\begin{align*}
  \frac{v_{ri}}{z_{ri}} = 
  1 - \sum_{\substack{s\in\mathcal{R}\\j\in\mathcal{I}_0}}
  C_{sj,ri}
  ~,
\end{align*}
that, in turn, is a consequence of \cref{eq:balance-app}. What
\cref{eq:attr-income-app} tells us is that the total value of the income for which
each region's consumption is responsible is just that region's spending. And,
this is the only consistent attribution if global income is to equal global
spending. Leontiefian assumptions \ref{itm:L1-app} and \ref{itm:L2-app} suppose that
environmental impacts may be attributed along the same lines as monetary flows.
The fact that $\hat{A}_{ri,s}$ accurately attributes income to spending is a
reflection of these assumptions and plays a central role in the appearance of
majorization.

\subsection{Environmentally extended tables}
\label{sec:env}

As mentioned already, we wish to explore the use of MRIO tables to attribute
the impacts of economic activities to the demand sectors that stimulate them.
Impacts themselves are often accounted for by environmentally extending the
input-output table. A MRIO table with environmental extension is an
\emph{environmentally-extended} MRIO (EE-MRIO) table.

In the multiregional setting, an environmental extension is a family of block
vectors $\{\mathbf{e}^{(\alpha)}\}$, indexed by impact $\alpha$, with the form
$e^{(\alpha)}_{ri}$, $r\in\mathcal{R}$ and
$i\in\mathcal{I}_0$. The quantity $e^{(\alpha)}_{ri}$ gives the total impact of
the activity in sector $i$ and region $r$ during the same time period as the
other input-output matrices. For instance, if $\alpha$ corresponds to
greenhouse gas emissions, then $e^{(\alpha)}_{ri}$ is the quantity of
greenhouse gases emitted measured in carbon dioxide equivalents. The regional
impacts are then given by $\hat{e}^{(\alpha)}_r =\sum_{i}e^{(\alpha)}_{ri}$ and
the total impacts are denoted $E^{(\alpha)} = \sum_r \hat{e}^{(\alpha)}_r$.

Given impact $\alpha$, we define the \emph{$\alpha$-intensity}
$f^{(\alpha)}_{ri}$ of sector $(r,i)$ as the ratio of environmental impact to
economic impact.  While there are many different definitions, for our needs we
define it as:
\begin{align*}
  f^{(\alpha)}_{ri} := \frac{e^{(\alpha)}_{ri}/v_{ri}}{E^{(\alpha)}/Y}
  ~.
\end{align*}
That is, we take the ratio of the impact $e^{(\alpha)}_{ri}$ 
to the \emph{value-added} 
$v_{ri}$ by
sector $(r,i)$. This directly relates the emission at a given stage of
production to the value added to the region by that production. (Not counted in
this is the economic input from other industries, as this value has already
been counted as value-added in another industry.) To remove dependence on the
units used, we normalize the ratio by comparing it to the total ratio between
emissions and income.

We can similarly define the \emph{regional intensity} by
\begin{align}
\label{eq:reg-intensity-app}
  \hat{f}^{(\alpha)}_{r} & :=
  \frac{\hat{e}^{(\alpha)}_{r}/\hat{y}_{r}}{E^{(\alpha)}/Y} \\
  & = \sum_{i\in\mathcal{I}_0}\frac{v_{ri}}{\hat{y}_r}f^{(\alpha)}_{ri}
  \nonumber
  ~.
\end{align}
As \cref{eq:reg-intensity-app} indicates, this can be conceptualized either as a
ratio of totals or the regional average of sectoral intensities.

Impacts happen at the point of production, but this
activity meets a demand somewhere potentially geographically
distant. Impacts may be thought of as becoming {\em embodied}
in their product, which travels from production to demand
\cite{wiedmann_editorial_2009}.
EE-MRIO tables have been frequently used to compute
the flow of embodied impacts from production to final demand.
While impact at the point of production is described
the impacts vector $\mathbf{\hat{e}}^{(\alpha)}$,
the embodied impacts attributed to each region $r$
are given by the attribution vector:
\begin{align}
\label{eq:attr-impact-app}
  \hat{\mathbf{a}}^{(\alpha)}:= \mathbf{e}^{(\alpha)}\mathbf{\hat{A}}
  ~.
\end{align}

In theory, the quantity $\hat{a}^{(\alpha)}_r$ describes the total impact,
originating anywhere in the global economy, required to meet the demands of
region $r$. We again emphasize
that this depends on the assumptions
\ref{itm:L1-app} and \ref{itm:L2-app}.
For now, it is sufficient to appreciate that these are the standard
calculations employed in Leontief analysis of EE-MRIO tables.

Embodied flows resulting from Leontief analysis are frequently quantified by
one or both of the following proxy measures \cite{erb_embodied_2009,
davis_consumption-based_2010, bergmann_bound_2013, moran_does_2013,
yu_tele-connecting_2013, alsamawi_employment_2014, simas_bad_2014,
dorninger_can_2015, liu_carbon_2015, bergmann_land_2016, oita_substantial_2016}:
\begin{enumerate}
      \setlength{\topsep}{-5pt}
      \setlength{\itemsep}{-5pt}
      \setlength{\parsep}{-5pt}
\item The net export $\boldsymbol{\xi}^{(\alpha)}=(\xi^{(\alpha)}_r)$ 
of attributed impact, as a share of total global impact:
  \begin{align*}
    \xi^{(\alpha)}_r 
    = \frac{\hat{e}^{(\alpha)}_r - \hat{a}^{(\alpha)}_r}{\sum_r \hat{e}^{(\alpha)}_r} ~.
  \end{align*}
\item Or, the ratio $\boldsymbol{\rho}^{(\alpha)}=(\rho^{(\alpha)}_r)$ 
  of attributed impact to direct impact:
  \begin{align*}
    \rho^{(\alpha)}_r
    = \frac{\hat{a}^{(\alpha)}_r}{\hat{e}^{(\alpha)}_r}
  ~.
  \end{align*}
\end{enumerate}
Naturally, these are closely related, as they ultimately express the
relationship between the relative sizes of produced impacts and consumed
impacts.

\section{Null model construction}
\label{app:null-model}

The structure of the null model is primarily founded on a distinction among the
technical coefficients, between those that are truly ``technical'' (in that
they depend directly on production techniques) and those that are ``social''
(in that they may depend on exogenous social parameters, such as acceptable
wage levels, trade agreements, and consumer ethics). Technical constraints, as
calculated by the GTAP 8 dataset, are emulated in the null model. Social
constraints, on the other hand, are randomized entirely.

To begin, we define the regional technical coefficients
$\hat{C}_{i,sj}$ and global technical coefficients $\tilde{C}_{i,j}$ to be:
\begin{align*}
  \hat{C}_{i,sj} = \frac{\sum_r Z_{ri,sj}}
  {\sum_{r,i} Z_{ri,sj}} ~, \quad
  \tilde{C}_{i,j} = \frac{\sum_{r,s} Z_{ri,sj}}
  {\sum_{r,i,s} Z_{ri,sj}}
  ~.
\end{align*}
The first characterizes the input requirements of regional industries relative
to other industries. Note that the social choice of from which regions to
acquire inputs is not described by this matrix. The second $\tilde{C}_{i,j}$
averages these industrial requirements over the entire globe.

Our null model is then constructed as follows:
\begin{enumerate}
      \setlength{\topsep}{-5pt}
      \setlength{\itemsep}{-5pt}
      \setlength{\parsep}{-5pt}
\item The global technical coefficients $\tilde{C}_{ij}^{(\mathrm{GTAP})}$
	from GTAP 8 are used as a baseline to generate regional technical
	coefficients. Namely, for each region $r$ and industry $j$ the technical
	coefficients $\hat{C}_{i,rj}$ are drawn as:
	\begin{align*}
		\hat{C}_{i,rj} \sim \mathrm{Dir}\left(\alpha_i = \zeta_{C,i}\hat{C}_{i,rj}\right)
	~.
	\end{align*}
	This ensures that the technical composition of local industries is similar
	to that found in the GTAP 8 database. The vector parameter $\zeta_{C,i}$
  controls the
	degree of similarity.
\item The regional spending distribution $s_{a,r}$, consumption coefficients
	$c_{i,ra}$, import coefficients $M_{i,sj}$, regional supply coefficients
	$R_{r,si}$, value-added coefficients $U_{ri}$, and factor coefficients
	$F_{u,ri}$ are all drawn from uniform Dirichlets:
	\begin{align}
		\begin{split}
		s_{a,r} &\sim \mathrm{Dir}\left(\alpha_a = 1\right)\\
		c_{i,ra} &\sim \mathrm{Dir}\left(\alpha_i = 1\right)\\
		(M_{i,sj},1-M_{i,sj}) &\sim \mathrm{Dir}\left(\alpha_1=1,\alpha_2=1\right)\\
		R_{r,si} &\sim \mathrm{Dir}\left(\alpha_r=1\right)\\
		(U_{ri},1-U_{ri}) &\sim \mathrm{Dir}\left(\alpha_1=1,\alpha_2=1\right)\\
		F_{u,ri} &\sim \mathrm{Dir}\left(\alpha_u=1\right)
    \end{split}
    ~.
\end{align}
	In order, these describe: the distribution of regional spending among the
	final demand sectors; the distribution of spending by each final demand
	sector among its products of consumption; the proportion by which a given
	industrial sector will import a particular input instead of source it
	domestically; the regional probability of importing an input from another
	specific region; the proportion of capital outlay towards factors by a
	given industrial sector; and the distribution of that capital among the
	factors. These describe socially-determined relations between the regions
	and sectors. These are the relationships we seek to randomize in the null
	model.
\item The full technical coefficients are constructed as:
	\begin{align*}
		C_{ri,sj} = \begin{cases}
		(1-U_{ri})(1-M_{i,sj})\hat{C}_{i,sj} & r = s\\
		(1-U_{ri})R_{r,si}M_{i,sj}\hat{C}_{i,sj} & r \neq s
		\end{cases}
	    ~.
	\end{align*}
	From these we define the matrix:
	\begin{align*}
		K_{r,s} := \sum_{i,j,a}U_{ri}\left(\mathbf{I}-\mathbf{C}\right)^{-1}_{ri,sj}
		c_{j,sa}s_{a,s}
		~.
	\end{align*}
	This matrix describes the likelihood that money originating in a final demand sector in region $s$ ends up in a value-added sector in region $r$.
	Now, given any particular global spending distribution $\mathbf{\hat{x}}$,
	the regional incomes are determined by
	$\hat{\mathbf{y}}=\mathbf{K}\mathbf{\hat{x}}$. We calculate the eigenvector
	$\boldsymbol{\pi}$ such that $\mathbf{K}\boldsymbol{\pi}=\boldsymbol{\pi}$.
	(Its existence is guaranteed by the fact that $\mathbf{L}$ is stochastic in
	its left index.) If spending matched this eigenvector, then no region would
	hold a trade deficit or surplus. All trade would be equally balanced. The
	spending in our model is drawn from a Dirichlet as:
	\begin{align*}
		\hat{x}_r \sim \mathrm{Dir}(\alpha_r = \zeta_X \pi_r)
	    ~.
	\end{align*}
	 Thus, the parameter $\zeta_X$ controls the scale of trade imbalance.
\item The multiregional input-output table can now be constructed as:
	\begin{align}
    \begin{split}
      D_{ri,ra} &= c_{i,ra}s_{a,r}\hat{x}_r\\
      Z_{ri,sj} &= 
      \sum_{\substack{t\in\mathcal{R}\\
      k\in\mathcal{I}_0\\
      a\in\mathcal{D}_0}}
      C_{ri,sj}
      \left(\mathbf{I}-
      \mathbf{C}\right)^{-1}_{sj,tk}
    D_{tk,ta}\\
    V_{ru,ri} &=
    \sum_{\substack{s\in\mathcal{R}\\
      j\in\mathcal{I}_0\\
      a\in\mathcal{D}_0}}
      F_{u,ri}U_{ri}
      \left(\mathbf{I}-
      \mathbf{C}\right)^{-1}_{ri,sj}
    D_{sj,sa}
    \end{split}
    ~.
	\end{align}
From $Z_{ri,sj}$ we compute the sector activities
	$z_{ri}=\sum_{s,j}Z_{ri,sj}$ and finally the attribution matrix:
	\begin{align*}
		A_{ri,sa} = \frac{\left[
		\left(\mathbf{I}-\mathbf{C}\right)^{-1}
		\mathbf{D}
		\right]_{ri,sa}}{z_{ri}}
	~.
	\end{align*}
\end{enumerate}

The parameters will be generally set at baseline values $\bar{\zeta}_{C,i}$ and
$\bar{\zeta}_X$, determined by:
\begin{align}
  \begin{split}
    \bar{\zeta}_{C,i}^{-1} &= \sum_{
      \substack{r\in\mathcal{R}\\j\in\mathcal{I}_0}}
    \frac{v_{ri}^{(\mathrm{GTAP})}}
    {\sum_s v^{(\mathrm{GTAP})}_{si}}
    \hat{C}^{(\mathrm{GTAP})}_{j,ri}\log\left(
      \frac{\hat{C}^{(\mathrm{GTAP})}_{j,ri}}
      {\tilde{C}^{(\mathrm{GTAP})}_{j,i}}
    \right)\\
    \bar{\zeta}_{X}^{-1} &= \sum_{r\in\mathcal{R}}
    \hat{x}^{(\mathrm{GTAP})}_r \log\left(
      \frac{\hat{x}_r^{(\mathrm{GTAP})}}
      {\pi_r^{(\mathrm{GTAP})}}
    \right)
  \end{split}
  ~.
\end{align}
This uses the Kullback-Liebler divergence \cite{Cove06a}
as a proxy for the degree of
difference between various empirical distributions. This sets the baseline for
similar variations within the null model.
\end{document}